\documentclass[twocolumn,superscriptaddress,showpacs]{revtex4}

\usepackage{amsbsy}
\usepackage{amsfonts}
\usepackage{amssymb}
\usepackage{amsmath}
\usepackage{graphicx,color}
\usepackage{graphicx}

\begin{document}


\title{Physically Realizable Entanglement by Local Continuous Measurements}

\author{Eduardo Mascarenhas}
\affiliation{Departamento de F\'isica, Universidade Federal de Minas Gerais, Belo Horizonte, Caixa Postal 702, 30123-970, MG, Brazil}
\affiliation{Centre for Quantum Technologies, National University of Singapore, Singapore}

\author{Daniel Cavalcanti}
\affiliation{Centre for Quantum Technologies, National University of Singapore, Singapore}

\author{Vlatko Vedral}
\affiliation{Centre for Quantum Technologies, National University of Singapore, Singapore}
\affiliation{Department of Physics, National University of Singapore, Singapore}
\affiliation{Department of Physics, University of Oxford, Clarendon Laboratory, Oxford, OX1 3PU, UK}

\author{Marcelo Fran\c{c}a Santos}
\affiliation{Departamento de F\'isica, Universidade Federal de Minas Gerais, Belo Horizonte, Caixa Postal 702, 30123-970, MG, Brazil}
\affiliation{Centre for Quantum Technologies, National University of Singapore, Singapore}

\begin{abstract}
Quantum systems prepared in pure states evolve into mixtures under environmental action. Physically realizable ensembles are the pure state decompositions of those mixtures that can be generated in time through continuous measurements of the environment. Here, we define physically realizable entanglement as the average entanglement over realizable ensembles. We optimize the measurement strategy to maximize and minimize this quantity through local observations on the independent environments that cause two qubits to disentangle in time. We then compare it with the entanglement bounds for the unmonitored system. For some relevant noise sources the maximum realizable entanglement coincides with the upper bound,  establishing the scheme as an alternative to locally protect entanglement. However, for local strategies, the lower bound of the unmonitored system is not reached.
\end{abstract}
\maketitle

Decoherence is the process in which the exchange of information between a quantum system and an external environment continuously downgrades quantum properties of the former~\cite{dec}. This dynamics turns initially pure entangled states into mixed less entangled ones, destroying the capacity of the system for quantum applications. In this picture, the description of the properties of the system in time results from ignoring the information that leaks into the environment and only considering the general statistical characteristics of the reservoir. If, however, one is able to recover some of this information by, for example, continuously observing changes into the environment, the time evolution of the system undergoes a different dynamics which is conditioned on the results of these observations. Such conditional evolutions are usually referred to as quantum trajectories~\cite{dec} because if the system is initially prepared in a pure state, the sequence of measurements performed on the environment determines a respective sequence of other pure states for the system. Each trajectory corresponds to one realization of the experiment and the open system evolution is recovered when averaging over all possible trajectories. Such trajectories have already been observed both in massive particles and light~\cite{jump1}. 

There are infinitely many different ways to observe a physical environment, also referred to as unravellings. Each one of them defines a different set of experimentally realizable trajectories. However, even though at any given time the incoherent sum over all possible trajectories reproduces the quantum state of the unmonitored system, that does not mean that all its pure state decompositions are available, since not all ensembles can be achieved through this continuous monitoring process~\cite{PRE}. The fact that one can reconstruct the state allows one to recover the average value of any observable in time by averaging over the trajectories. However, when it comes to obtaining the entanglement of the system, the restrictions imposed by the unraveled time evolution may be too strong~\cite{Andre1}. 

In this paper, we investigate equivalent local measurement strategies over the independent reservoirs that cause two entangled qubits to lose coherence and entanglement in time. The restriction on the ensembles that can be continuously generated in time becomes clear when we compare the minimum and maximum average entanglement~\cite{Carmichael} resulting from locally realizable unravellings to the extremes over all possible decompositions of a given density matrix, respectively the entanglement of formation $E_F$~\cite{Bennett} and the entanglement of assistance $E_A$~\cite{Assistance}. The locality restriction applied means that the measurements performed over independent reservoirs should, themselves, be independent as well, or at most classically correlated. This local condition is essential whenever the decohering qubits are distributed over different and distinct channels, like in quantum teleportation, communication or likewise. 

We analyze both dephasing and spontaneous emission as the decoherence sources. For dephasing, we show that it is not only possible to continuously reach the entanglement of assistance of the system but the protocol to do so does not even require any classical communication between the observers, as long as local phase feedback is allowed. On the other hand, we also show that the entanglement of formation can never be achieved through local observations, even with the addition of classical noise to the measurement process~\cite{Diff1}, already making explicit the impact of the missing decompositions on the realizable entanglement. This impact becomes even more explicit for dissipative reservoirs where $E_A$ can only be achieved for some particular initial states, and, once again, $E_F$ can never be achieved through local observations. This means that not only some decompositions are not produceable through local continuous monitoring of the environment, but, more important, neither is the entanglement of $\rho(t)$. This also means that the independent monitoring of local reservoirs is indeed a good strategy to preserve entanglement for quantum communication and teleportation protocols, specially in quantum feedback schemes~\cite{Feedback}. 

The physical system we describe is that of two non-interacting qubits initially prepared in a pure state $\sigma=|\sigma\rangle \langle \sigma|$ and coupled to independent reservoirs in the Bohr-Markovian approximations~\cite{dec}. Their unmonitored time evolution is determined by a master equation of the type $\dot{\rho}=\mathcal{L_A}\rho+\mathcal{L_B}\rho$ where
\begin{equation}
\mathcal{L}\rho=\sum_{i} \frac{\gamma_i}{2}(2\Gamma_i \rho \Gamma_i^\dagger - \{\Gamma_i^\dagger \Gamma_i, \rho\}),
\end{equation}
and the set $\{\Gamma_i\}$ represents the subsystem-reservoir couplings. $\rho(t)$ can be formally obtained by integrating this equation in time. It also corresponds to a completely positive map in the Kraus form~\cite{Disc} acting on the initial state,
\begin{equation}
\rho=\textbf{K}\sigma\textbf{K}^{\dagger}=\sum_i K_i\sigma K_i^{\dagger},
\label{CPmap}
\end{equation}
with $\textbf{K}$ representing a vector of Kraus operators $\textbf{K}= \left[ 
    K_0 , K_1 , \cdots , K_n\right]$, and $\textbf{K}^{\dagger}= \left[K_0^{\dagger} , K_1^{\dagger}, \cdots, K_n^{\dagger}\right]^{\mathrm{T}}$. The $i$th Kraus operator can be identified as a measurement operator such that probability is conserved, meaning  $\sum_i K_i^{\dagger}K_i=\openone$, and the measurement process is characterized by a non negative distribution $\langle K_i^{\dagger}K_i\rangle\ge 0$. The $\textbf{K}$ unravelling gives output states representing a decomposition of the system state 
$\rho=\Psi\Psi^{\dagger}$, with $\textbf{K}|\sigma\rangle=\Psi$, and $\Psi_i=K_i|\sigma\rangle=|\widetilde{\psi}_i\rangle$ (the tilde endicates a unnormalized state). 

This is one of infinitely many possible decompositions of $\rho$ and any other decomposition $\rho= \Phi\Phi^{\dagger}$ can be obtained by a unitary rearrangement of $\Psi$, $\Phi^{\mathrm{T}}=\mathcal{U}\Psi^{\mathrm{T}}$, such that $
\rho=\Phi\Phi^{\dagger}=\Psi \mathcal{U}^{\mathrm{T}}\mathcal{U}^{\ast}\Psi^{\dagger}=\Psi\Psi^{\dagger}$ where $\mathcal{U}$ is a unitary matrix respecting $(\mathcal{U}^{\dagger}\mathcal{U})^{\mathrm{T}}=\mathcal{U}^{\mathrm{T}}\mathcal{U}^{\ast}=\openone$. In~\cite{HJW}, the authors show a one to one relation between unravellings and pure state ensembles by demonstrating that this same unitary freedom applies to the unravelling choice, or in other words, that the same unitary takes the unravelling $\textbf{K}$ that generates $\Psi$ to another unravelling $\textbf{G}$ that generates $\Phi$, with $\textbf{G}^{\mathrm{T}}=\mathcal{U}\textbf{K}^{\mathrm{T}}$:
\begin{equation}
\rho=\textbf{K}\sigma\textbf{K}^{\dagger}=\textbf{K}\mathcal{U}^{\mathrm{T}}\sigma\mathcal{U}^{\ast}\textbf{K}^{\dagger}=\textbf{G}\sigma\textbf{G}^{\dagger}.
\label{G}
\end{equation}
Their result shows that in principle all possible ensembles that decompose the decohered state at a given time $t$ can be obtained through measurements on a purification of the state. These include, in particular, the decompositions that handle the maximum and minimum average entanglement, respectively $E_A$ and $E_F$. 

However, the measurement apparatuses that lead to some decompositions are not compatible with a continuous and local measurement process of the environments, or, as defined in~\cite{PRE}, the corresponding ensembles are not physically realizable through the continuous and local monitoring of the independent reservoirs. 

If one is to monitor the reservoirs continuously, then the evolution of the system respects a new equation where at each time step the system is projected onto a new state. Assuming that the system is initially prepared in a pure state, the evolution is then given by a sequence of pure states $\{|\sigma_0\rangle,|\sigma_{dt}\rangle,...,|\sigma_{ndt}\rangle\}$ which are called quantum trajectories of the system~\cite{dec}. This process is also described by a CP-map in the form of (\ref{CPmap}) but now there is an extra restriction imposed over the different possible unravellings $\textbf{K}$. Physically, this is related to the fact that some $\textbf{K}$'s represent operations that combine measurements performed on the reservoir on different times, while the Markovian approximation that leads to the master equation imposes reservoirs with no memory which, under constant monitoring, must produce measurement outcomes that are not correlated in time. Furthermore, by imposing local observation of the reservoirs that act in each qubit, one also creates an extra restriction on the class of unitaries $\mathcal{U}$ that combine the different unravelling operations to unitaries of the type $\mathcal{U_A}\otimes \mathcal{U_B}$. 

Note that, in the limit of continuous monitoring there is yet another class of possible decompositions which is associated to the addition of classical noise to the measurement records (as in homodyne detection, for example~\cite{Wiseman2}). In this case, the decomposition reads $\textbf{G}_{\pm\ast}^{\mathrm{T}}=\frac{1}{\sqrt{2}}[\mathbf{\Omega}\pm\mathcal{U}\textbf{K}_{\ast}^{\mathrm{T}}]$, where $\ast$ indicates the absence of $K_0$ and $G_0$, and $G_0=\openone-\frac{1}{2}\sum_{i=1}[G_{i+}^{\dagger}G_{i+}+G_{i-}^{\dagger}G_{i-}]$. The vector $\mathbf{\Omega}$ (whose elements are complex constants) represents the external entity that adds the extra noise. When $|\mathbf{\Omega}|$ tends to infinity we reach the quantum state diffusion limit~\cite{Diff1}. When restricted to $\mathcal{U}=\mathcal{U_A}\otimes \mathcal{U_B}$ this decomposition always decreases the average entanglement as should be expected given that the added noise blurs the outcomes of the measurements on the reservoirs. This fact is used in~\cite{Andre1} to minimize the average concurrence.

We show bellow that the local and continuous restrictions made on the monitoring protocols of the reservoirs impose limitations on the minimum and maximum average entanglement  $\overline{E}$ attainable from realizable ensembles. Here, $\overline{E}(\Phi)=\sum_i p_i E(|\phi_i\rangle)$, where the entanglement of realizable state $\{|\phi_i\rangle\}$, that happens with probability $p_i=\langle\widetilde{\phi}_i|\widetilde{\phi}_i\rangle$, is obtained by calculating the Von Neumann entropy $S(\rho)=-\mathrm{tr}\{\rho \log{\rho}\}$ on the reduced density matrix over either one of the qubits, $E(|\phi_i\rangle)=S(\mathrm{tr}_{\mathrm{B}}\{|\phi_i\rangle\langle \phi_i|\})$.

\begin{figure}[h]
\includegraphics[width=9cm]{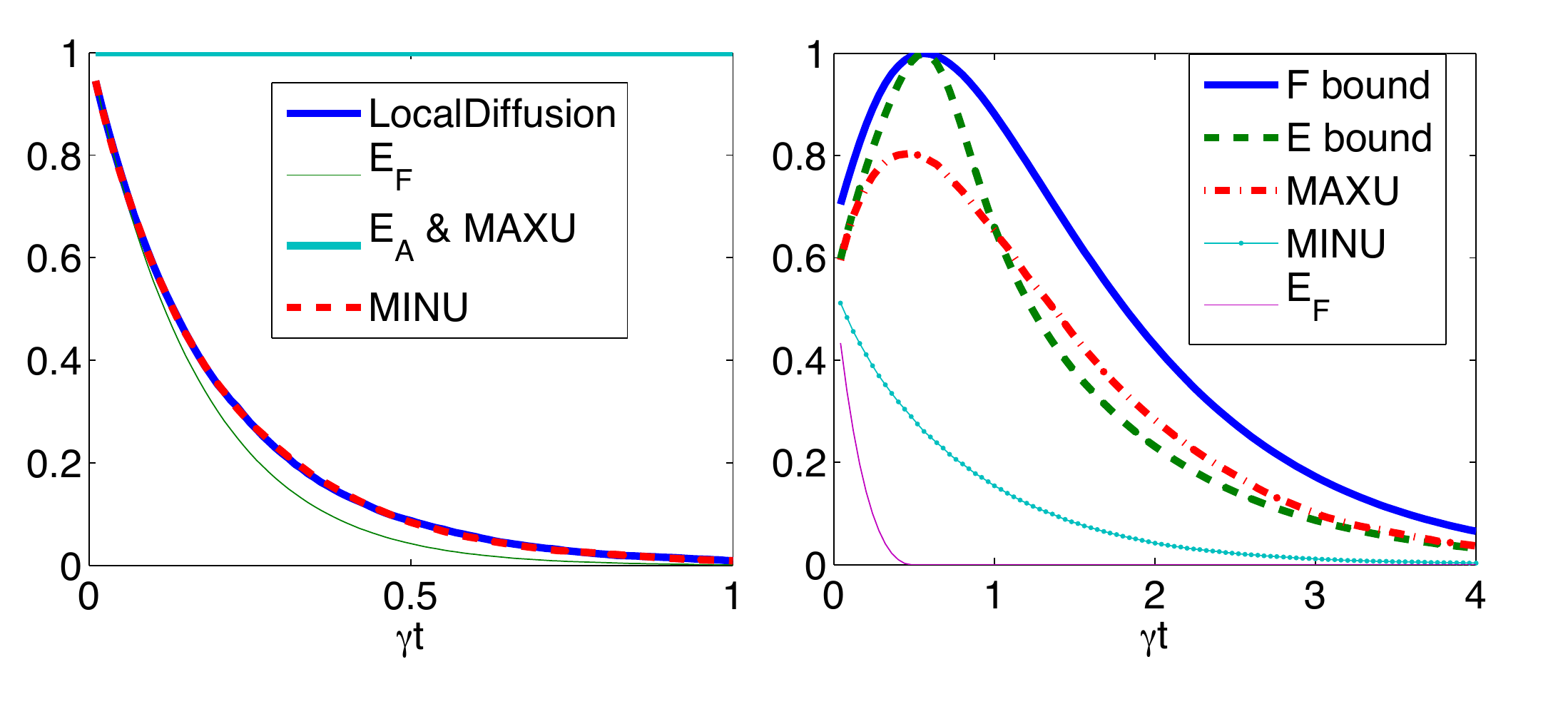}
\caption{(Left) Dephasing reservoir for a Bell state: there is a gap between the average entanglement unraveled by local quantum state diffusion and the entanglement of formation $E_F$. The entanglement of assistance $E_A$ and the maximum unraveled entanglement (MAXU) coincide. (Right) Dissipative reservoir for a state presenting sudden death of entanglement: quantum state diffusion still produces the minimum average entanglement which never goes to zero. In this case, there is a gap between both $E_F$ and $E_A$, and the respective minimum and maximum locally realizable entanglement. $E_A$ is in between the fidelity bound (F) and the eigenstate bound (E).}\label{F}
\end{figure}

Let us first consider the example of both qubits initially prepared in a maximally entangled state (for example one of $|\Phi_{\pm}\rangle=\frac{1}{\sqrt{2}}(|00\rangle\pm|11\rangle)$ and subjected to independent dephasing reservoirs represented by Kraus operators given by  $A_0=(1-\frac{\gamma dt}{2})\openone$ and $A_1=Z\sqrt{\gamma dt}$ in the short time expansion (up to first order in $dt$). Here $Z$ is the usual Pauli matrix, $\gamma$ is the system-environment coupling and the same holds for observer B. This particular unravelling ($\mathcal{U}_{\mathrm{A(B)}}=\openone$) already gives the $E_A$ of the system in each time interval $dt$~\cite{Dominique}. Dephasing implies that the system stochastically switches between $|\Phi_{+}\rangle$ and $|\Phi_-\rangle$ and by monitoring the reservoir one knows if the phase flip (the jump) has occurred or not. This result can still be classically communicated between the parts which then completely preserves the entanglement of the system. If, however, the observer is capable of phase flipping its local qubit, then not even the communication is required and the protocol becomes entirely local, similar to the one presented in ~\cite{Nos} for spontaneous emission. This also means that in this case $E_A[\rho(t)]$ corresponds to a locally realizable ensemble. Changing the unravelling also changes the decompositions and $\overline{E}$, however the $E_F$ of the unmonitored system is unreachable as shown in Fig~\ref{F} for the optimized local unravelling that minimizes $\overline{E}$.


Turning to dissipative environments, the $\mathcal{U}=\openone$ unravelling features Kraus operators of the type $A_0=\openone-\frac{\gamma}{2}L^{\dagger}L dt$ and $A_1=L \sqrt{\gamma dt}$, with $L=|0\rangle\langle1|$. For initial states in the subspace $\{|01\rangle,|10\rangle\}$ like, for example, $|\Psi_+\rangle=(|01\rangle+|10\rangle)/\sqrt{2}$, the entanglement of assistance of $\rho(t)$ can once again be exactly obtained and coincides with the fidelity bound~\cite{Assistance}. This derives from the particular time evolution of the system which leads to a mixture of a maximally entangled state and a separable one, $\rho(t)=p(t)|\Psi_+\rangle\langle \Psi_+|+[1-p(t)]|00\rangle\langle00|$, and $E_A[\rho(t)]$ is given by $p(t)=e^{-\gamma t}$. This maximum is exactly the one obtained by the $\mathcal{U}_{A(B)}=\openone$ unravelling~\cite{Nos,Dominique}, which means that once again, local strategies allow for the maximum extraction of entanglement from the system. In this case, classical communication is required given that a single decay completely destroys the entanglement of the system. Also note that, in this case, it is particularly important to use $E_F$ as a lower bound for the average entanglement since the Wootters concurrence, for example, still gives the same average values for all unravellings, as shown in~\cite{Andre1,Andre2}. 

From the point of view of the detection of unrealizable ensembles, the most interesting example, however, appears when we analyze initial states that present finite time disentanglement such as $|\sigma\rangle=\frac{1}{\sqrt{8}}(|00\rangle+\sqrt{7}|11\rangle)$~\cite{eu}. In this case, neither extremes of the entanglement of the system are achievable through local measurements on the independent reservoirs as shown in Fig. 1b. For the Entanglement of assistance, this derives immediately from the fact that local strategies cannot increase the average entanglement while $E_A$ is above the initial entanglement for times of the order of $1/\gamma$. The most significant difference, however, is between the realizable entanglement and $E_F[\rho(t)]$ because the local monitoring of the reservoirs always preserve some entanglement in the system while for the unmonitored evolution entanglement disappears in a finite time~\cite{eu}.

The persisting gap between $\overline{E}$ and $E_F$ necessarily means that ignoring the information that goes to the reservoir (master equation evolution of $\rho$) is different from continuously acquiring some of this information and then erasing it. The entropy of the measurement outcomes of a general unravelling is given by $S(\textbf{G})=-\sum_i p_i\log{(p_i)}$, with $p_i=\langle G_i^{\dagger}G_i\rangle$ being the probability of the $i$th outcome. The information acquired from the reservoir is maximum if the entropy goes to zero, and minimum if the entropy is maximum. For example, for a single reservoir, the $\mathcal{U}=\openone$ unravelling in the short time expansion is given by an operator $K_1$ proportional to $\sqrt{dt}$ (jump operator) and its complementary $K_0=\openone -\frac{1}{2}K^{\dagger}_1K_1$. The entropy, given by $S(\textbf{K})=- p_1\log{(p_1)}-(1-p_1)\log(1-p_1)$, with $p_1\propto dt$, tends to zero for small time steps $dt\rightarrow 0$, maximizing then the information. 

This unravelling locally produces the entanglement of assistance both for dephasing and spontaneous emission in the $\{|01\rangle,|10\rangle\}$ subspace. In both cases, the most probable event (no jump) is correlated to preserving entanglement in the system. In fact, for dephasing, both click and no click produce equally entangled states, hence the $E_A$ is locally realizable. Whereas, for the decay, no click still preserves entanglement, however, a click in the reservoir kills entanglement completely, hence $E_A$ decays. On the other hand, any other unravelling that combines the jump and no jump operators, necessarily increases the entropy, decreases the extracted information and mixes the above mentioned correlations. In the worst case scenario of a $50-50$ superposition ($\mathcal{U}=
\left[
\begin{array}{cc}
  1/\sqrt{2}   \quad  -1/\sqrt{2}   \\
  1/\sqrt{2} \quad   1/ \sqrt{2}   \\
\end{array}
\right]$) all the information is lost and the average entanglement is minimized. For dephasing, it means that an event (either click or no click) will be equally correlated to both $|\Phi_{\pm}\rangle$ states, while for the $\{|01\rangle,|10\rangle\}$ decay, it will be equally  correlated to the initial state and to $|00\rangle$. These limits establish the local unravelling bounds and any other local unravelling (more general $\mathcal{U}$'s) will produce average entanglement in between them. 

A similar analysis can be carried out in the diffusive limit. The measurement operators can be described by
$D_{\textbf{J}}=K_0+\textbf{G}_{\ast}\textbf{J}^{\dagger}$,
with $\textbf{G}_{\ast}=\textbf{K}_{\ast}\mathcal{U}^{\mathrm{T}}$ and $\textbf{J}dt=\langle \textbf{G}_{\ast}^{\mathrm{T}}+\textbf{G}_{\ast}^{\dagger}\rangle^{\mathrm{T}} dt+ d\textbf{W}$, such that $dW_i$ are infinitesimal Wiener increments (gaussian distributed variables of variance $dt$) and $d\textbf{W}^{\mathrm{T}}d\textbf{W}=\openone dt$. 
The short time expansion of the map that reproduces the density matrix is recovered with
\begin{equation}\rho(dt)=\int d\mu(\textbf{J}) D_{\textbf{J}}\sigma D_{\textbf{J}}^{\dagger},\end{equation}with $d\mu(\textbf{J})$ being the normalized gaussian measure over the measurement outcomes. In this case, the addition of classical noise,$dW_i$, blurs the outcomes and destroys the acquisition of information from the reservoir, which leads the diffusive limit also to the minimum local average entanglement, as we have shown in Fig.~\ref{F} of the paper. Once again, all the information is erased and the realizable entanglement reaches its minimum but not the $E_F[\rho(t)]$.

Before concluding let us comment on the possible changes related to lifting the local observation restriction. That means allowing for operations that coherently superpose the results observed in each independent reservoir. In Fig. 1, we show that this strategy may also increase the average entanglement of the system. This happens to be the case for initial states presenting finite time disentanglement for spontaneous emission. In this case, both the no jump trajectory and the superposed jumps are correlated to increasing entanglement for the first part of the evolution (times of the order of $1/\gamma$). While the no jump trajectory performs the optimum singlet conversion protocol~\cite{Vidal,Nos}, correlated jumps of the form  $G_{\pm}=\frac{1}{\sqrt{2}}[A_1\otimes\openone\pm \openone \otimes B_1]$ produce Bell states for both qubits, increasing then, the average entanglement of the system. This strategy is still not enough to reach the $E_A$ of the unmonitored evolution, but for certain times it redefines the lower bound for this quantity. In~\cite{Andre2} the authors use an equivalent strategy plus classical noise to reach the other extreme, continuously reproducing the concurrence of the unmonitored system.

We have studied  the entanglement that can be produced in quantum open systems when the environments coupled to the system are continuously monitored.  We have shown a gap between the locally realizable entanglement and the extremes entanglement of assistance and formation of the unmonitored system. We have also related the minimum attainable entanglement to the erasure of the information provided by the readout of the environment. On its turn, the above mentioned gap means that erasing this information is fundamentally different from not collecting it. 
This fact is better exemplified when the system is initially prepared in a pure entangled state and presents finite time disentanglement under dissipation, in which case it is impossible to locally unravel an ensemble with zero average entanglement at finite time. There is always at least one entangled trajectory which means that the average entanglement can only go to zero asymptotically in time. This fact in turn provides a method to protect entanglement and avoid entanglement sudden death. 
\acknowledgements
The authors would like to thank A.R.R. Carvalho for useful discussions. This work was financially supported by the National Research Foundation and the Ministry of Education of Singapore. E.M. would also like to thank CNPq for financial support.


\begin{thebibliography}{UEE}

\bibitem{dec}H. P. Breuer and F. Petruccione, The Theory Of Open Quantum Systems, Oxford University Press, Oxford, 2002.
\bibitem{jump1} S. Peil, and G. Gabrielse, Phys. Rev. Lett. \textbf{83}, 1287 (1999); W. Nagourney, J. Sandberg, and H. Dehmelt, Phys. Rev. Lett. \textbf{56}, 2797 (1986); T. Sauter, W. Neuhauser, R. Blatt, and P.E. Toschek, Phys. Rev. Lett. \textbf{57}, 1696 (1986); J.C. Bergquist, R.G. Hulet, W.M. Itano, and D.J. Wineland, Phys. Rev. Lett. \textbf{57}, 1699 (1986);
T. Basch\'e, S. Kummer, and C. Brauchle, Nature \textbf{373}, 132 (1995);  S. Gleyzes, \text{et. al.} Nature \textbf{446}, 297 (2007).
\bibitem{PRE} H.M. Wiseman and J.A. Vaccaro, Phys. Rev. Lett. \textbf{87}, 240402 (2001).
\bibitem{Andre1}  A.R.R. Carvalho, M. Busse, O. Brodier, C. Viviescas, \& A. Buchleitner, Phys. Rev. Lett. \textbf{98}, 190501 (2007).
\bibitem{Carmichael} H. Nha, and H.J. Carmichael, Phy. Rev. Lett. \textbf{93}, 120408 (2004).
\bibitem{Bennett} C.H. Bennett, D. P. DiVincenzo, J.A. Smolin, and W.K. Wootters, Phys. Rev. A \textbf{54}, 3824 (1996); W.K. Wootters, Phys. Rev. Lett. \textbf{80}, 2245 (1998).
\bibitem{Assistance} D.P. DiVincenzo, \textit{et. al.}, "The entanglement of assistance", in Lecture Notes in Computer Science \textbf{1509} (Springer-Verlag, Berlin, 1999), pp. 247; O. Cohen, Phys. Rev. Lett. \textbf{80}, 2493 (1998).
\bibitem{Diff1} N. Gisin, Phys. Rev. Lett. \textbf{52}, 1657 (1984); H. M. Wiseman and G. J. Milburn, Phys. Rev. A \textbf{47}, 1652 (1993)
\bibitem{Feedback} S. Mancini, and H.M. Wiseman, Phys. Rev. A \textbf{75}, 012330 (2007); N. Yamamoto, \textit{et. al.}, Phys. Rev. A. \textbf{78}, 042339 (2008); A.R.R. Carvalho, \textit{et. al.}, Phys. Rev. A, \textbf{78} 012334 (2008); C. Hill and J. Ralph, Phys. Rev. A \textbf{77}, 014305 (2008).
\bibitem{Disc} A. Shabani and D.A. Lidar, Phys. Rev. Lett. \textbf{102}, 100402 (2009).
\bibitem{HJW} L.P. Hughston, R. Josza, and W.K. Wootters, Phys. Lett. A \textbf{183}, 14 (1993).
\bibitem{Wiseman2} H. M. Wiseman and G. J. Milburn, Phys. Rev. A 47, 642 (1993).
\bibitem{Dominique}  After completion of this work we became aware that a similar result for the maximization of average entanglement was derived independently by S. Vogelsberger and D. Spehner,  arXiv:1006.1317.
\bibitem{Nos} E. Mascarenhas, B. Marques, D. Cavalcanti, M. T. Cunha, and M. F. Santos, Phys. Rev. A \textbf{81}, 032310 (2010).
\bibitem{Andre2} C. Viviescas, \textit{et.al.} arXiv:1006.1452 (2010)
\bibitem{eu} M. F. Santos, P. Milman, L. Davidovich and N. Zagury, Phys. Rev. A 73, 040305(R) (2006).
\bibitem{Vidal} G. Vidal, Phys. Rev. Lett. \textbf{83}, 1046 (1999).





\end{thebibliography}
\end{document}